\documentclass[allclo]{FBSart}
\usepackage{amsfonts}
\usepackage{amssymb}
\usepackage{epsfig}

\newcommand{\ga}{\alpha}
\newcommand{\gb}{\beta}
\newcommand{\gc}{\gamma}
\newcommand{\gd}{\delta}
\newcommand{\gve}{\varepsilon}
\newcommand{\gl}{\lambda}
\newcommand{\gs}{\sigma}
\newcommand{\gS}{\Sigma}

\title{Three- and Four-body correlations in nuclear matter}
\author{M.~Beyer\thanks{\textit{E-mail address:} 
 michael.beyer@physik.uni-rostock.de}}
\institute{Fachbereich Physik,
 Universit\"at Rostock, 18051 Rostock, Germany}

\runningauthor{M.~Beyer}
\runningtitle{Three- and Four-body correlations}
\begin{document}

\maketitle
\begin{abstract}
  Few-nucleon correlations in nuclear matter at finite densities and
  temperatures are explored. Using the Dyson equation approach
  leads to effective few-body equations that include self energy
  corrections and Pauli blocking factors in a systematic way. Examples
  given are the nucleon deuteron in-medium reaction rates, few-body
  bound states including the $\ga$-particle, and $\ga$-particle
  condensation.
\end{abstract}

\section{Introduction}

Strongly correlated many-particle systems provide an exciting field
for applications of few-body methods. Examples are nuclear matter,
quark matter (quark gluon plasma), ionic plasmas, among others.  In an
equilibrium situation at finite temperatures and densities one usually
introduces a mean field to describe the gross features of
many-particle systems. However, many exciting phenomena, such as the
formation of clusters below a certain density (Mott density) and the
appearance of superconductivity below a critical temperature, cannot
be described in a framework of noninteracting quasiparticles (ideal
system). Within a quantum statistical approach it is possible to go
beyond such a picture and consider the residual interactions between
the quasiparticles. A well known equation, e.g., to study two-body
correlations is known as Bethe-Goldstone equation and/or
Feynman-Galitskii equation, depending on details~\cite{fet71}.  To go
beyond two-body correlations is desirable for many reasons.  Among
them are: {\it i)} For a microscopic description of the heavy ion
collision, three-body reaction rates are an important input into the
collision integral.  Their medium dependence has hardly been
studied~\cite{Beyer:1996rx,Beyer:1997sf,Kuhrts:2000jz,Kuhrts:2001zs}.
{\it ii)} Bound states are effected by the density and temperature of
the medium and the binding energy may become zero leading to the Mott
effect~\cite{Beyer:1999zx,Beyer:1999tm,Beyer:2000ds,dan91,Dan92}.
{\it iii)} Since the $\alpha$-particle is the strongest bound nucleus
it should be relevant for the equation of state of nuclear matter. It
might induce $\alpha$-condensation and quartetting which would be a
different state of matter besides superfluidity induced by pairing or
pair condensation~\cite{rop98}.  {\it iv)} The three-quark system is
particularly interesting because it is a reasonable model of baryons
and useful to study the phase transition between the quark phase and
the hadronic phase of nuclear matter~\cite{Beyer:2001bc,mat01}.  {\it
  v)} The three-body input into the two-body t-matrix implied by the
equations hierarchy has not been studied and might, e.g., affect the
critical temperature.

The first three issues will be addressed in the following, {\it iv)}
is given to some detail in Ref.~\cite{mat01} of this issue, and {\it
  v)} is left for future investigations and is relevant, e.g., for the
question of a possible color superconducting phase.

\section{Theoretical tools}
We utilize the Dyson equations to tackle the many-particle problem. An
review is given in Ref.~\cite{duk98}. This enables us to decouple the
hierarchy of equations finally deriving effective in-medium few-body
equations. The many-particle Hamiltonian is given by
\begin{eqnarray}
H =\sum_{1} H_0(1)a_1^\dagger a_{1}+\;\frac{1}{2}\;
\sum_{12 1'2'}V_2(12,1'2')\;a^\dagger_1  a^\dagger_2  a_{2'} a_{1'}
\end{eqnarray}
where $H_0$ is the single particle energy and $V_2$ a generic two-body
potential. The chronological equal time cluster Green function for a
given number of particles is defined by
\begin{eqnarray} 
{\cal G}_{\alpha\beta}^{\tau-\tau'}&=&-i\langle T_\tau A_\ga(\tau)
A_\gb^\dagger(\tau')\rangle,\qquad{\langle\dots\rangle
={\rm Tr}\{\rho_0\dots\}}
\end{eqnarray}
where $A_\ga=a_1a_2a_3\dots a_n$, e.g., are Heisenberg operators and
$\rho_0$ the equilibrium density operator. The respective Dyson
equation for the cluster is
\begin{eqnarray} 
i\frac{\partial}{\partial\tau}\; {\cal G}^{\tau-\tau'}_{\ga\gb}
&=&\delta(\tau-\tau')  \underbrace{\langle [A_\ga,A_\gb^\dagger]_\pm\rangle}
_{=:{\cal N}^\tau_{\ga\gb}} 
+ \sum_{\gc}\int d\bar \tau\; {{\cal M}^{\tau-\bar \tau}_{\ga\gc}}
\;{\cal G}^{\bar \tau-\tau'}_{\gc\gb}.
\label{eqn:dyson}
\end{eqnarray}
The mass matrix that appears in (\ref{eqn:dyson}) is given by
\begin{eqnarray} 
 {\cal M}^{\tau-\tau'}_{\ga\gb}
&=& \underbrace{\delta(\tau-\tau') {\cal M}^{\tau}_{\ga\gb,0}}_{
  \mbox{\rm cluster mean field}} 
 +
 \underbrace{{\cal M}^{\tau-\tau'}_{\ga\gb,\rm irr.}}_{
 \mbox{\rm   retardation}}\label{eqn:mass}\\
( {\cal M}{\cal N})^\tau_{\ga\gb ,0}&=
&\langle [[A_\ga,H](\tau),A_\gb^\dagger(\tau)]\rangle\\
({\cal MN})^{\tau-\tau'}_{{\rm irr.},\alpha\beta}&=&
\sum_\gamma
\langle T_\tau [A_\alpha,H]_\tau,[A^\dagger_\gb,H]_{\tau'}]
\rangle_{{\rm irreducible}}.
\end{eqnarray} 
More details are given in \cite{duk98}.  The equation
(\ref{eqn:dyson}) is expressed in momentum space and the time
component as Matsubara frequency $\tau \rightarrow z_\gl$
that is analytically continued into the complex plane, $
z_\gl\rightarrow z$, for a textbook treatment see \cite{fet71}.  To
arrive at suitable calculable expressions the following approximations
are utilized: {\it i)} Only the cluster mean field contribution to the
kernel (\ref{eqn:mass}) is used; {\it ii)} The density operator $\rho$
is evaluated for an uncorrelated medium. This way the equations
hierarchy is decoupled and effective few-body equations that describe
few-body correlations including medium effects have been derived.  For
$A_\ga=a_1$, the one-particle Green functions is
\begin{equation}
G(z)=(z-\varepsilon_1)^{-1}
\end{equation}
where the quasi-particle self energy is
\begin{equation}
\varepsilon_1 = \frac{k^2_1}{2m_1}+\sum_{2}V_2(12,\widetilde{12})f_2
\simeq \frac{k^2_1}{2m_1^{\rm eff}}+\gS^{\rm HF}(0).
\end{equation}
The last equation introduces the effective mass that is a valid
concept for the rather low densities considered here and $\mu^{\rm
  eff}\equiv \mu - \gS^{\rm HF}(0)$.  The Fermi function $f_i\equiv
f(\varepsilon_i)$ for the $i$-th particle is given by
\begin{equation}
f(\varepsilon_i) = \frac{1}{e^{\gb(\varepsilon_i - \mu)}+1}.
\end{equation}
The resolvent $G_0$ for $n$ noninteracting quasiparticles is
\begin{equation}
G_0(z) = (z-  H_0)^{-1}
N \equiv R_0(z) N,\qquad H_0 = \sum_{i=1}^n \varepsilon_i
\end{equation}
where $G_0$, $H_0$, and $N$ are formally matrices in $n$ particle
space. The Pauli-blocking for $n$-particles is
\begin{equation}
N=\bar f_1\bar f_2 \dots \bar f_n
\pm f_1f_2\dots f_n,\qquad\bar f=1-f
\end{equation}
where the upper sign is for Fermi-type and the lower for Bose type
clusters. Note that $NR_0=R_0N$. Straight forward evaluation of
(\ref{eqn:dyson}) using the Wick theorem leads also to the full
resolvent $G(z)$. For each number of particles in the cluster the resolvents
have the same formal structure and can be written in a convenient way
close to the one for the isolated system, viz.
\begin{equation}
G(z)=(z-H_0-V)^{-1}{N}
\equiv R(z){N}, \qquad 
V\equiv \sum_{\mathrm{pairs}\;\ga} N_2^{\ga}V_2^{\ga}.
\end{equation}
Note that $V^\dagger\neq V$ and $R(z)N\neq NR(z)$. To be specific,
for an interaction in pair $\ga=(12)$ the effective potential
reads 
\begin{equation}
\langle 123\dots|{N_2^{(12)}}V_2^{(12)}|1'2'3'\dots\rangle = 
{(\bar f_1\bar f_2 - f_1f_2)}
V_2(12,1'2')\gd_{33'}\dots
\end{equation}
For further use in the Alt Grassberger Sandhas (AGS) equations~\cite{alt67}
we give also the channel resolvent
\begin{equation}
G_\ga(z)=(z-H_0-  N_2^{\ga}V_2^{\ga})^{-1} N
\equiv R_\ga(z) N.
\end{equation}
For the two-body case as well as for a two-body subsystem embedded in
the $n$-body cluster the standard definition of the $t$ matrix leads
to the Feynman-Galitskii equation for finite temperature and
densities~\cite{fet71},
\begin{equation}
T_2^\ga(z) =   V_2^\ga + 
 V_2^\ga  N^\ga_2 R_0(z)  T_2^\ga(z).
\label{eqn:T2}
\end{equation}
Introducing the AGS transition operator via
\begin{equation}
R(z)=\delta_{\alpha \beta}R_\alpha(z)+
R_\alpha(z){U_{\alpha\beta}(z)} R_\beta(z)
\end{equation}
the effective inhomogeneous in-medium AGS equation reads
\begin{equation}
U_{\alpha\beta}(z)= (1-{\delta}_{\alpha\beta})R^{-1}_0(z)+
\sum_{\gamma\neq \alpha}
{N^\gc_2}
T_2^\gamma(z)R_0(z)
U_{\gamma\beta}(z).
\label{eqn:T3}
\end{equation}
The homogeneous in-medium AGS equation uses the form factors defined by
\begin{equation}
|F_\gb\rangle\equiv\sum_\gc\bar\gd_{\gb\gc} { N_2^\gc}  V_2^\gc 
|\psi_{B_3}\rangle
\end{equation}
to calculate the bound state $\psi_{B_3}$. Because of the non-symmetric
form of the potential the equation for the form factors and the dual
are different
\begin{eqnarray}
|F_\ga\rangle
&=&\sum_\gb \bar\gd_{\ga\gb}  
{N_2^\gb} T_2^\gb(B_3) R_{0}(B_3)|F_\gb\rangle\\
|\tilde F_\ga\rangle
&=&\sum_\gb \bar\gd_{\ga\gb} T_2^\gb(B_3) { N_2^\gb} R_{0}(B_3)
|\tilde F_\gb\rangle.
\end{eqnarray}
Finally, the four-body bound state is described by
\begin{equation}
|{\cal F}^\gs_\gb\rangle=\sum_{\tau\gc} \bar\gd_{\gs\tau}
\underbrace{U^\tau_{\gb\gc}(B_4)}_{\mbox{3-body}}  R_0(B_4) { N_2^\gc} 
\underbrace{T_2^\gc(B_4)}_{\mbox{2-body}} R_0(B_4) |{\cal F}^\tau_\gc\rangle,
\qquad\ga\subset\gs,\gc\subset\tau.
\end{equation}
where $\gs,\tau$ denote the four-body partitions. The two-body input
is given in (\ref{eqn:T2}) and the three-body input by (\ref{eqn:T3}),
both medium dependent.

\section{Results}
\subsection{Reaction rates}
An experiment to explore the equation of state of nuclear matter is
heavy ion collisions at various energies. Here we focus on
intermediate to low scattering energies and compare results to a
recent experiment $^{129}$Xe+$^{119}$Sn at 50 MeV/A by the INDRA
collaboration~\cite{INDRA}. A microscopic approach to tackle the heavy
ion collision is given by the Boltzmann equation for different particle
distributions $F\equiv(f_p,f_n,f_d,f_t,f_h)$
up to $h$ and $t$~\cite{dan91,Dan92},
\begin{equation}
\frac{\partial}{\partial t} F(t)+\{\gve,F(t)\}=
{\cal K}^{\rm in}_i[F(t)]\,(F(t) \pm 1)
-{\cal K}^{\rm out}_i[F(t)]\, F(t)
\label{eqn:Boltz} 
\end{equation} 
where $\gve$ is the mean field energy and ${\cal K}$ denotes the
respective collision integrals that include, e.g.,
the one for deuteron loss $Nd\rightarrow NNN$,
\begin{eqnarray}
{\cal K}^{\rm out}_d(P,t)&=&
\int d^3k\int d^3k_1d^3k_2d^3k_3\;
\underbrace{|\langle k_1k_2k_3|U_0|kP\rangle|^2_{dN\rightarrow pnN}}
_{\mbox{medium dependent!}}\nonumber\\&&
\qquad\times
\bar f_N(k_1,t)\bar f_N(k_2,t)\bar f_N(k_3,t)f_N(k,t)+\dots\label{eqn:react2}
\end{eqnarray}
where $\bar f \equiv 1-f$. A solution is given via a Boltzmann Uehling
Uhlenbeck (BUU) simulation ~\cite{dan91,Dan92}. As indicated in
(\ref{eqn:react2}) the reaction rate is in principle medium dependent.
However, previously this medium dependence has been neglected.  Within
linear response theory for infinite nuclear matter the use of
in-medium rates leads to faster time scales for the deuteron life time
and the chemical relaxation time as has been shown in detail in
Refs.~\cite{Beyer:1997sf,Kuhrts:2000jz}.

\begin{figure}[htb]
\begin{minipage}{0.49\textwidth}
\epsfig{figure=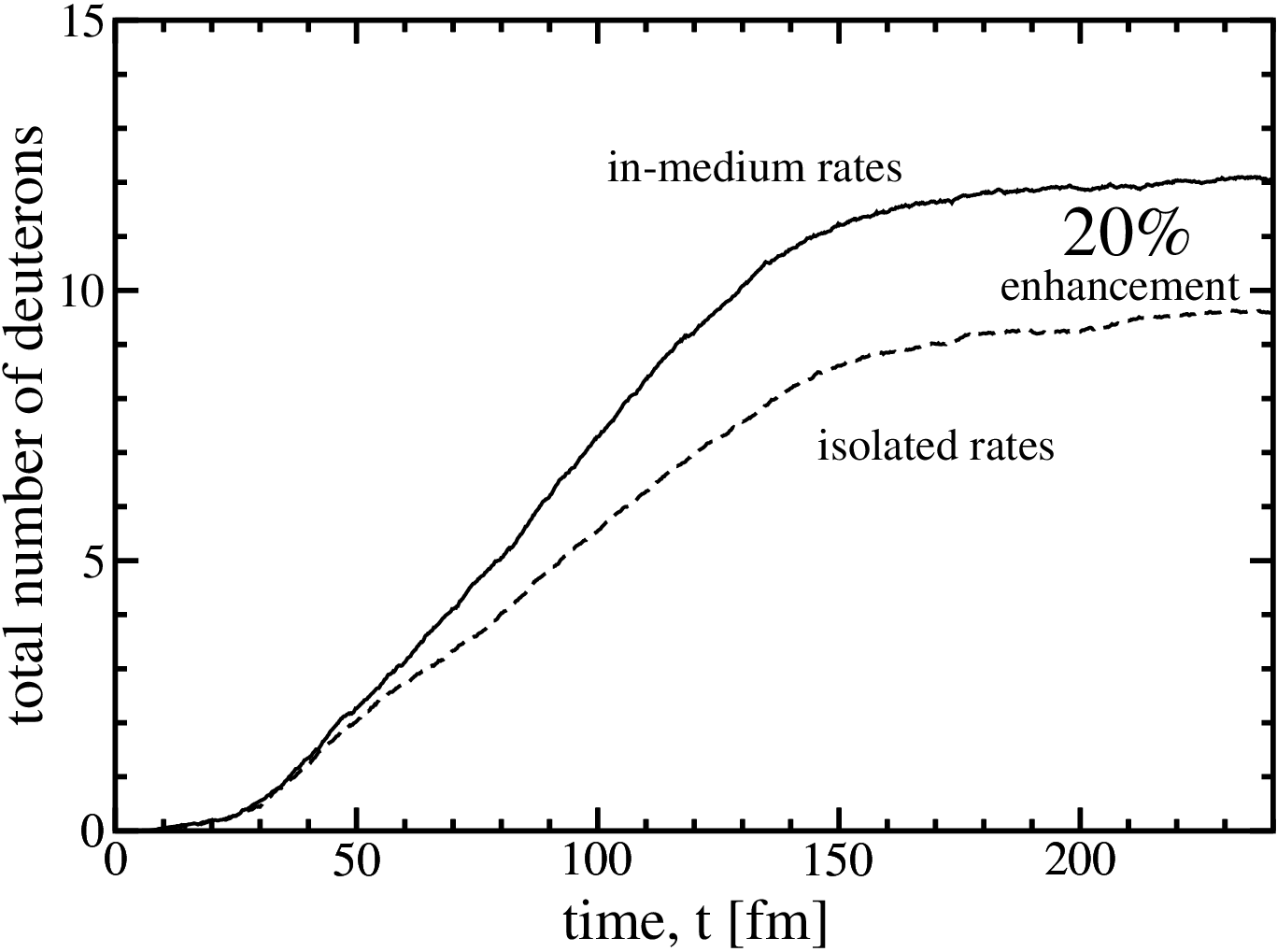,width=0.9\textwidth}
\caption{\label{fig:PLT} 
  BUU simulation of the deuteron formation during the central collision of
  $^{129}$Xe+$^{119}$Sn at 50 MeV/A.}
\end{minipage}\hfill
\begin{minipage}{0.49\textwidth}
  \epsfig{figure=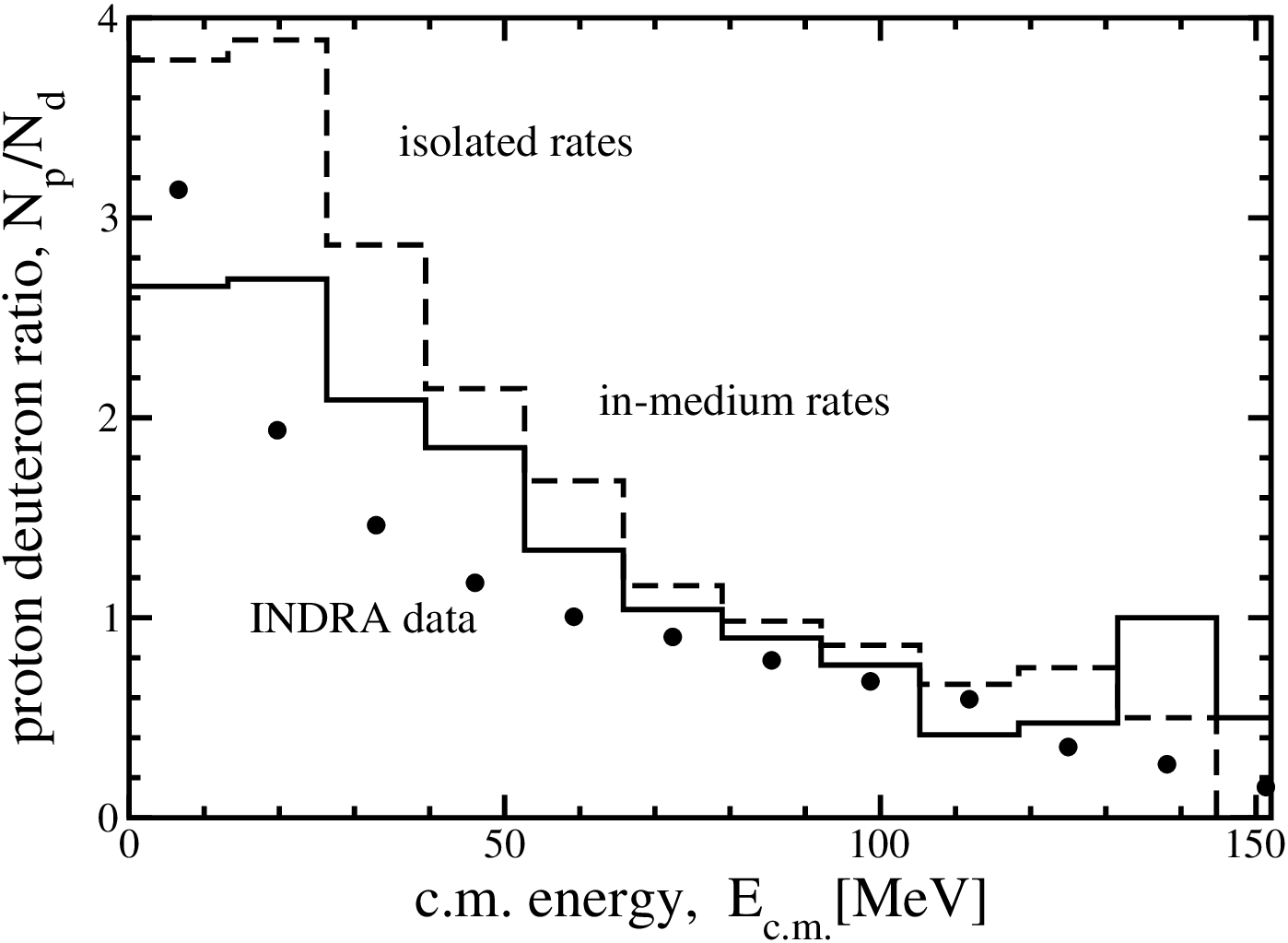,width=0.9\textwidth}
\caption{\label{fig:pd} 
  Ratio of proton to deuteron numbers as a function of c.m. energy. The
  experimental data are from the INDRA collaboration.}
\end{minipage}
\end{figure}
Now we use the in-medium AGS equations (\ref{eqn:T3}) that reproduce
the experimental data in the limit of an isolated three-body system.
For details on the specific interaction model see
Ref.~\cite{Beyer:1996rx}. We investigate the influence of medium
dependent rates in the BUU simulation of the heavy ion collision as
compared to use of isolated (i.e. experimental) rates.
Figure~\ref{fig:PLT} shows that the net effect (gain-minus-loss,
eq.(\ref{eqn:react2})) of deuteron production becomes larger for the
use of in-medium rates (solid line) compared to using the isolated
rates (dashed line). The change is significant, however, a comparison
with experimental data is difficult since deuterons may also be
evaporating from larger clusters that has not been taken into account
in the present calculation so far. The ratio of protons to deuterons
may be better suited for a comparison to experiments that is shown in
Figure \ref{fig:pd}. The use of in-medium rates (solid line) lead to a
shape closer to the experimental data (dots) than the use of isolated
rates (dashed line).

\subsection{Bound states, Mott effect}
\begin{figure}[htb]
\begin{minipage}{0.49\textwidth}
\epsfig{figure=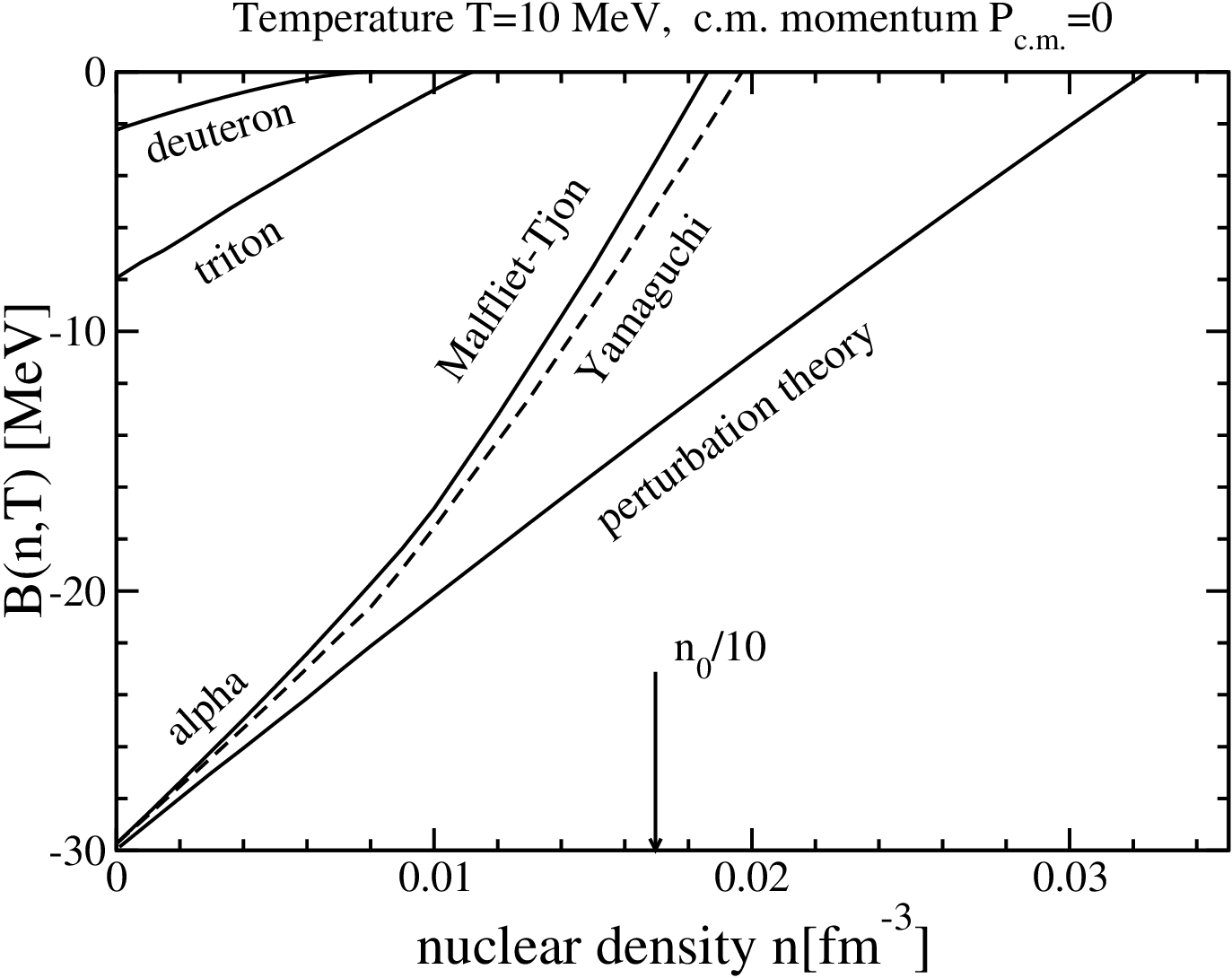,width=0.9\textwidth}
\caption{\label{fig:amottC} 
  Difference between the pole energy of the bound state and the
  continuum, $B(n,T)=E_{\rm pole}-E_{\rm cont}$.}
\end{minipage}\hfill
\begin{minipage}{0.49\textwidth}
\epsfig{figure=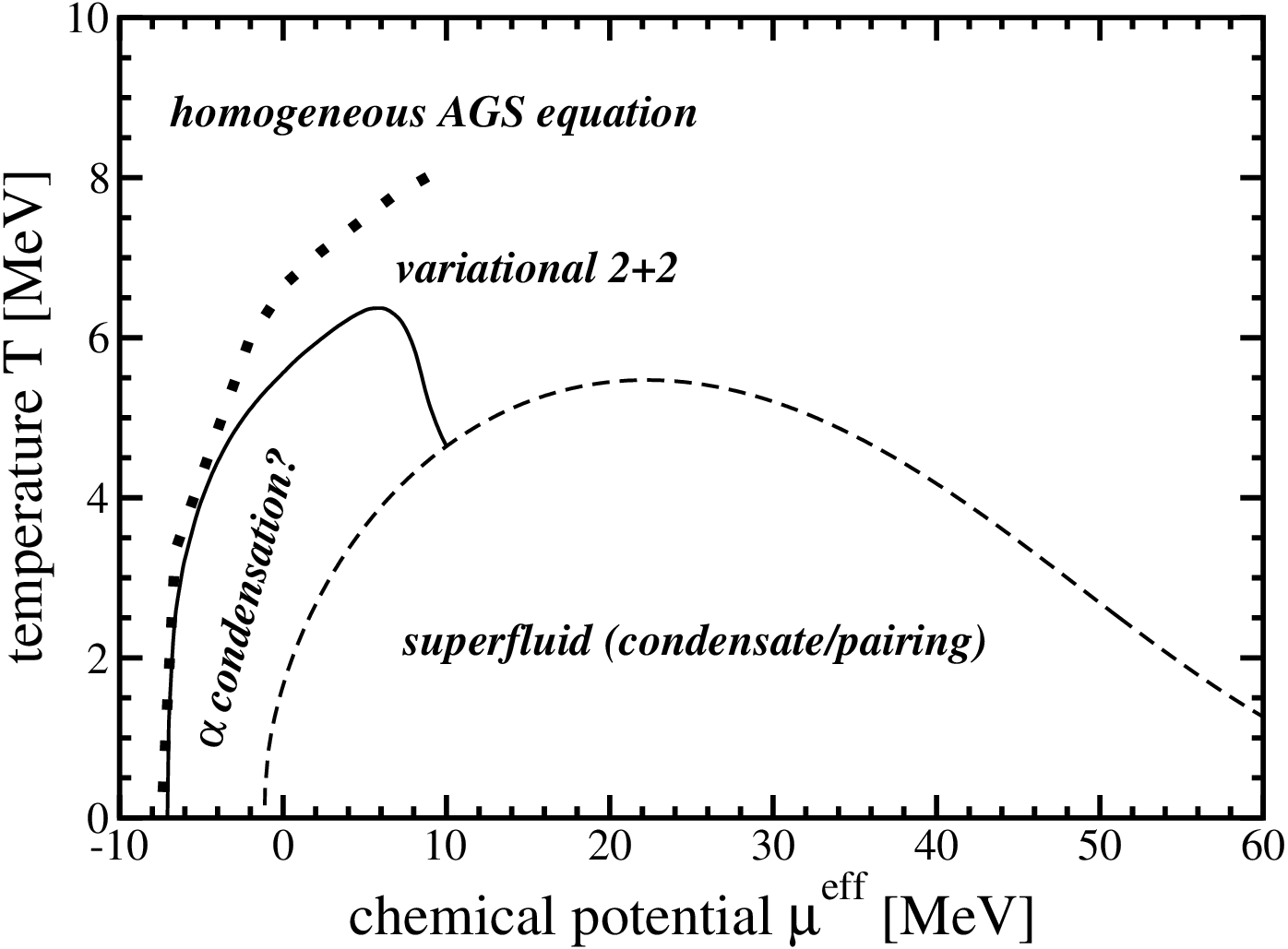,width=0.9\textwidth}
\caption{\label{fig:Tc} 
  Critical temperatures of condensation/pairing leading to superfluid
  nuclear matter. For an explanation see text.}
\end{minipage}
\end{figure}
In these calculation, besides the change of rates, also the Mott
effect has been taken into account. Figure \ref{fig:amottC} shows the
dependence of the binding energy for different clusters at a given
temperature of $T=10$ MeV and at rest in the medium. At first sight an
Efimov effect \cite{efi70} might be expected in the vicinity of the
Mott transition of the deuteron. However, two main reasons prevent the
Efimov effect to appear in a simple way: {\it i)} The deuteron binding
energy in the medium depends parametric (through the blocking factors)
on the deuteron momentum. Since the deuteron-like subsystem is not at
rest, in other words the effective strength of the potential that
enters into the three-body problem varies with momenta, a possible
Efimov effect is washed out. {\it ii)} The excited states that should
appear from the continuum (Efimov states) are as well blocked by the
medium.  This blocking may not be so strong as the ground states,
because the momentum distribution is peaked at higher momenta. As a
consequence only a careful quantitative analysis might answer the
question of Efimov states. On the other hand nuclear matter might not
be the best system to eventually observe such an effect.

In Figure \ref{fig:Tc} part of the phase diagram of nuclear matter is
shown. The lines indicate phase transitions. The critical temperatures
of condensation/pairing (dashed line, \cite{rop98}) leading to
superfluid nuclear matter are shown. The possible area of $\alpha$
condensation (solid line) as suggested by~\cite{rop98} is also given.
The latter is based on a variational calculation using the 2+2
component of the $\alpha$ particle to evaluate the condition for the
onset of superfluidity for the four-particle system
$B(T_c,\mu,P=0)=4\mu$.  The critical temperature found by solving the
homogeneous AGS equation for $\mu<0$ confirms the onset of $\alpha$
condensation even at higher values (dotted line). For $\mu>0$ the
condition $B=4\mu$ for the phase transition can also be fulfilled.
However, the homogeneous AGS equation cannot be used to investigate
the steep fall-off predicted in Ref.~\cite{rop98} because of continuum
poles that are not compensated by the blocking factors of the
potential as is the case for the two-body problem.  Whether the steep
fall-off is of physical origin or due to the use of a homogeneous
equation also for $\mu>0$ needs further investigation.

\begin{acknowledge}
Work supported by Deutsche Forschungsgemeinschaft.
\end{acknowledge}

\end{document}